# Defect-Mediated Relaxation in the Random Tiling Phase of a Binary Mixture: Birth, Death and Mobility of an Atomic Zipper

Elisabeth Tondl, Malcolm Ramsay, Peter Harrowell and Asaph Widmer-Cooper

*School of Chemistry, University of Sydney*

*Sydney NSW 2006 Australia*

**Abstract**

This paper describes the mechanism of defect-mediated relaxation in a dodecagonal square-triangle random tiling phase exhibited by a simulated binary mixture of soft discs in 2D. We examine the internal transitions within the elementary mobile defect (christened the 'zipper') that allow it to move, as well as the mechanisms by which the zipper is created and annihilated. The structural relaxation of the random tiling phase is quantified and we show that this relaxation is well described by a model based on the distribution of waiting times for each atom to be visited by the diffusing zipper. This system, representing one of the few instances where a well defined mobile defect is capable of structural relaxation, can provide a valuable test case for general theories of relaxation in complex and disordered materials.

# 1. INTRODUCTION

Given the large structural fluctuations present, it is not hard to imagine that relaxation in an amorphous material can proceed via the excitation of local reorganizations ('defects') that can then move the system between distinct configurations. The problem is that, in the absence of any explicit characterization of the localised reorganization, this proposal can only aspire to a fuzzy phenomenology. Crystals pose the converse problem. The identification of defects is straightforward but the point defects cannot relax the structure due to the high degree of global constraint imposed by the very structure that made the defects definable. This is the basic dilemma faced by efforts to describe amorphous relaxation. One resolution of this quandary is to find a condensed phase of sufficient order that defects could be explicitly identified and of sufficient *disorder* that the system has a degenerate manifold of distinct groundstates whose exploration would constitute structural relaxation. We propose that the dodecagonal random square-triangle phase exhibited by some binary mixtures in 2D is just such a phase. In this paper we report on the nature of mobile point defects in the random tiling (RT) phase of a 2D mixture, determine the mechanism and rate of mobility of these objects and demonstrate the relationship between their motion and structural relaxation. We make no claim that the mechanism of structural relaxation of the RT phase provides any general account of relaxation in amorphous states. Rather, our goal is to present as complete an account as possible of slow collective structural relaxation in a dense phase characterised by a degeneracy of groundstates. Such a system can provide a reference system against which more general treatments of amorphous relaxation can be tested.

**1.1 Models of Defect-Mediated Relaxation of Amorphous States**

Phenomenological models of amorphous relaxation based on defects have a long history. Over 50 years ago, Sivert Glarum [1] modelled the dielectric relaxation in isoamyl bromide as the result of diffusing defects. He demonstrated that the solution of the 1D version of this model produced a stretched exponential relaxation function similar to that observed experimentally. The extension of this model to 3D was considered by a number of workers [2] leading to Bordewijk's exact asymptotic solution [3] in 1975 that showed, on increasing the dimension, that the relaxation function had become a simple exponential. Stretched exponential relaxation in 3D was recovered by Schlesinger and Bendler [4] who considered the case where the defect's random walk was governed by a power law distribution of waiting times, as opposed to the exponential distribution of a Poisson process. A number of papers [5] have explored this model in detail.

The diffusing defect model made no assertion as to the nature of the defect itself. Early suggestions drew on the idea of 'free volume' and pictured the defect as a vacancy [6] or a pair of rotating particles [7]. Mooney [8] criticized these models in the context of stress relaxation, arguing that such objects were the wrong symmetry to couple to a shear stress. Granato [9,10] has argued that point defects in the form of a pair of atoms occupying a single lattice site can produce, at sufficient concentrations, liquid like-behaviour. This interstitial model has been extended to amorphous solids and liquids where the mechanical properties and energetics of the interstitial defects have been retained while the actual identity of the localised objects is left unspecified.

**1.2 Crystal Defects in 2D**

The properties and dynamics of point defects in the triangular crystal in 2D have been the subject of both experimental [11] and simulation studies [12-15]. Libal et al [12] injected either a vacancy or an interstitial into a simulated 2D crystal. They found for both cases that the defect could adopt a number of configurations characterised by aggregations of dislocations. Specifically, they identified two defects consisting of a pair of dislocations (one offset and one linear), a symmetric triplet and a ring of 4 dislocations. These compound defects matched those observed in 2D colloid suspensions [11]. The movement of the defect consisted of glides along crystallographic directions (via the dislocation pair configurations) interspersed with direction changes involving the high symmetry triplet structure. Subsequent simulations studies [15] in which defects were thermally excited have confirmed that the dominant mechanism of motion involved transitions between the two dislocation pair configurations.

## 1.3 The Square-Triangle Random Tiling

In a dodecagonal square-triangle tiling, the ratio of the number of equilateral triangles $n_{tr}$ to the number of squares $n_{sq}$ can be varied continuously. This tiling can be decorated with a binary mixture of $N_S$ small particles (placed at the centers of the squares) and $N_L$ large particles (places at the vertices), such that $N_S = n_{sq}$ and $N_L = n_{tr}/2 + n_{sq}$. The maximum configurational entropy of the random tiling is found at a ratio of squares to triangles $n_{sq} / n_{tr} = \sqrt{3} / 4$ [16], which corresponds to a mole fraction of small particle $x_S = N_S/(N_S + N_L) = 0.317$. (At this optimal ratio the squares and triangles contribute equally to the area of the tiling.) The entropy per vertex at this optimal ratio is 0.119 [17], which corresponds to an entropy of 0.082 per particle at the optimal mixture composition. If the existence of two possible configurations requires that the entropy equals *ln*2, then, on average, 8 to 9 particles are required to produce a 2-state configuration space. We shall refer to the phase formed at this composition as the RT phase. Despite the random character of the RT phase, it is not a glass. The transition between the RT phase and the liquid is first order. The structure of the RT phase is characterised by a global 12-fold orientational order and the resulting scattering pattern exhibits well defined peaks [16,18]. This RT phase is often referred to as a dodecagonal quasicrystal, although strictly only a subset of all possible random tilings at this composition have ideal quasicrystalline order as characterised by zero phason strain [16,18].

Random square-triangle tilings have been widely studied as a model for twelve-fold symmetric quasicrystals [18,19] and, experimentally, the structure of some quasicrystalline alloys [20], binary nano-particle superlattices [21,22] and soft matter assemblies [23,24] have been described that can be mapped onto square-triangle tilings. Simulations of particles interacting via attractive patches [25] have also resulted in square-triangle nets. These various manifestations of random tilings typically arise from particle interactions quite different from the case we study here and so the specific defects and their associated dynamics described in this paper will not transfer directly to other examples of RT phases.

To conclude this introduction of the square-triangle tiling, we shall consider its connection with glasses. A rapid quench of the binary liquid mixture we study here will produce an arrested liquid, i.e. a glass, which is quite distinct from the RT phase achieved through a first order freezing transition upon sufficiently slow cooling. A review [26] of phase transitions in metal alloys found that many alloys have amorphous to quasicrystal phase transitions upon de-vitrification (i.e. crystallization on heating), raising the possibility of some structural relationship between the amorphous and quasicrystalline states. In a study of the nucleation kinetics of Ti-Zr-Ni alloys [27], it was established that the free energy barrier to nucleation of the quasicrystal was smaller than that to the more stable cubic crystal structures. In this paper we make no claims regarding the structural connection between the glass and the RT phase. We simply note that the RT phase shares the following features with a glass: a groundstate degeneracy with an extensive $T = 0$ configurational entropy, a diversity of local coordination environments and an absence of regular layered structures. The degree to which our study of structural relaxation in the low temperature RT phase can shed light on structural relaxation in glasses depends on the physical significance of these common features in the relaxation of the amorphous system.

## 2. MODEL AND ALGORITHM

The binary soft disc model consists of particles interacting via purely repulsive potentials of the form $u_{ab}(r) = \varepsilon(\sigma_{ab} / r)^{12}$ where $a$, $b$ = S or L. The independent parameters for the model are $\sigma_{LL}$, $\sigma_{SL}$ and $x_S$, the fraction of small particles (where we have defined the unit length to be $\sigma_{SS}$). We shall fix $\sigma_{LL} = 1.4$ and $\sigma_{SL} = 1.0$ (significantly smaller than the additive value of $[\sigma_{LL} + \sigma_{SS}]/2 = 1.2$) [28]. The systems used in this paper consist of $N = 4896$ particles enclosed in a square box with periodic boundary conditions. All units quoted will be reduced so that $\sigma_{SS} = \varepsilon = m = 1.0$, where $m$ is the mass of both types of particle. Specifically, the reduced units of time and temperature are $\tau = \sigma_{SS}(m/\varepsilon)^{1/2}$ and $T = k_B T/\varepsilon$, respectively. The composition will be specified using $x_S = N_S/N$, the mole fraction of the smaller of the particles, and all results presented will be for $x_S = 0.317$.

Previously [28], we have described the sequence of low temperature phases observed for different choices of the length $\sigma_{SL}$. As $\sigma_{SL}$ is increased from 1.0 up to 1.3 we observed the formation of a glass with medium-range RT order and defects (at $\sigma_{SL} = 1.0$), a compound AB crystal with a unit cell containing 4 particles (at $\sigma_{SL} = 1.1$) , a glass (at $\sigma_{SL} = 1.2$) and phase separated crystals of pure S and L particles (at $\sigma_{SL} = 1.3$). Subsequent to the completion of most of this work, we also found that $\sigma_{SL} = 0.9899$, a small decrease from the value used here, significantly increases the stability of the RT phase, while not affecting most of the results presented here.

The molecular dynamics (MD) simulations were carried out at constant pressure ($P = 13.5$) and temperature using Nose-Hoover style non-Hamiltonian equations of motion as implemented in LAMMPS [29]. A defect-free RT configuration was constructed (by replicating the random tiling shown in Fig. 1 of Ref. 18 in a 2 by 2 pattern), and heated slowly in steps from $T = 0$ until melting occurred at $T = 0.49$. At $T = 0.45$ and above we observed the spontaneous formation of zipper defects. Statistical analysis was subsequently performed using additional trajectories, propagated at temperatures in the range $T = 0.40$-$0.48$, originating from configurations with isolated defects. To filter out the statistical noise caused by thermal fluctuations, we analysed the inherent structure motion along these trajectories by first minimizing the energy of all the configurations using the conjugate gradient method, i.e. all results reported were generated by analysing inherent structure configurations unless explicitly stated otherwise.

## 3. RESULTS

### 3.1 The Structure and Properties of Local Thermally Excited Defects in the RT Phase

The RT phase is constructed out of the four vertices shown in Fig. 1. All other vertices correspond to defects known as *disclinations* and are characterized by a s*trength n* [16] related to the sign and magnitude of the angular distortion required to fit an illegal number of squares and triangles about a vertex, specifically,

$$n = 3n_{sq} + 2n_{tr} - 12 \qquad\qquad (1)$$

Note that $n = 0$ for each of the four RT vertices in Fig. 1 and is positive if there is a nominal angular excess around the vertex and negative if there is a nominal angular deficit. Isolated disclinations are not observed at low temperatures. Instead, we find a +1 and a -1 disclination associated as a nearest neighbour pair. Such a pair of disclinations is referred to as a *dislocation*. In our simulations, we observe immobile and persistent (i.e. stable) defects consisting of two and three dislocations. (Isolated dislocations are not observed at low T.) The most common immobile defects, labelled static doublet (SD) and static triplet (ST), are depicted in Fig. 2.

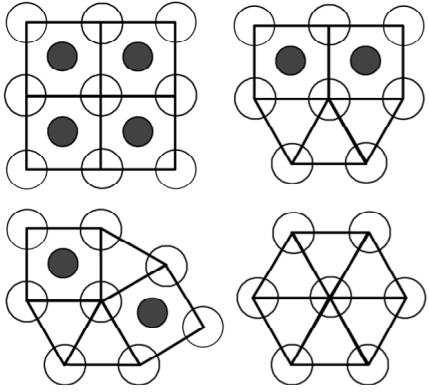

**Figure 1.** The four types of large particle vertices allowed in the ideal RT phase. The positions of the large and small particles has been indicated by open and filled circles, respectively.

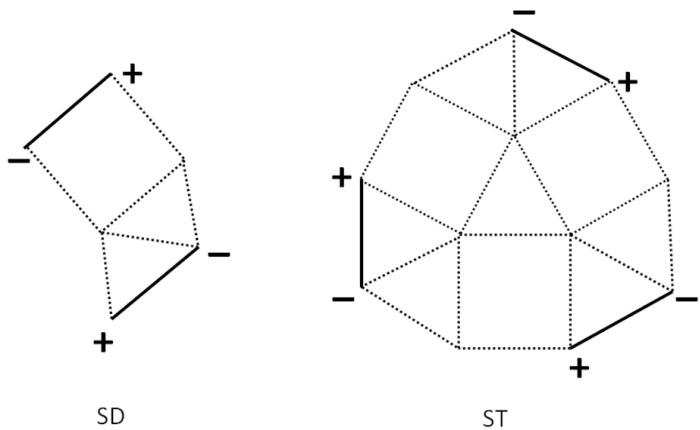

**Figure 2.** The static doublet (SD) and static triplet (ST) of dislocations. The strength of each disclination, as defined in Eq. 1, is indicated. The vertices indicate the position of large particles. Each square encloses a small particle which, for clarity, we do not show.

There also exists a mobile defect. The passage of this defect through the RT phase in the course of an individual MD trajectory is depicted in Fig. 3. A careful examination of Fig. 3 reveals that the topological changes associated with the passage of the defect are also accompanied by changes in the positions of the vertices themselves, i.e. structural relaxation. Previously, Leung et al [16] noted that "… the smallest subset of a square and triangle tiling that can be rearranged, leaving the rest fixed, is a chain of width two cells." This description certainly fits the behaviour represented in Fig. 3, however no details were provided in Ref. 16 concerning an explicit mechanism for the transformation between different RT configurations.  Oxborrow and Henley [18] later introduced a relaxation move for a tiling model in which a square and adjacent triangle were replaced by a triangle and two new thin tiles which could subsequently propagate until they recombined. The thin tiles do not correspond to any possible stable defect in the atomic mixture we consider in this paper, however the defect proposed in Ref. 18 is observed in a more elaborate atomic model for the RT phase described in Ref. 30.

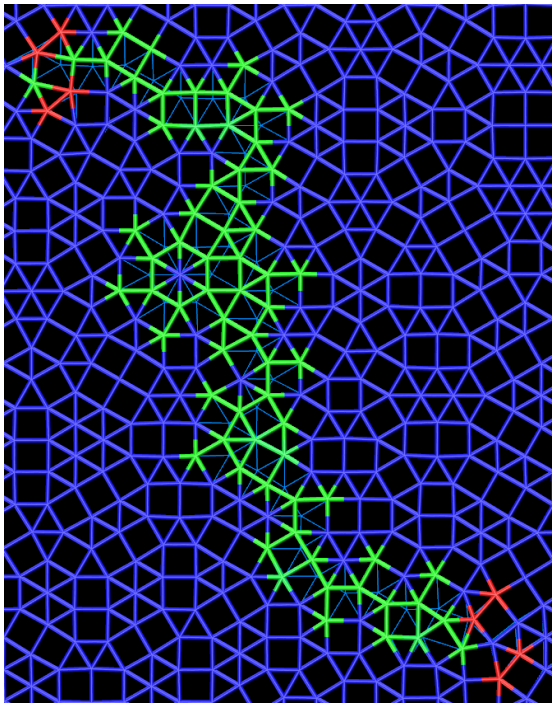

**Figure 3**. The passage of a mobile defect through the RT phase during an MD trajectory at $T = 0.45$. The initial configuration is shown as a purple net whose edges

link nearest neighbour large particles. Those large particles that have experienced a change of neighbours by the end of the trajectory, along with the edges of this final net, are indicated in green. The origin of the zipper is indicated by the SD defect (red) in the lower right corner and the final position of the mobile defect (red) is at the end of the line of topological change in the top left corner. A movie of the zipper motion is provided in the Supplementary Material [31].

The propagating defect observed in simulations is, like the static doublet, composed of two dislocations, but with a different set of mutual arrangements between which transformation can occur as shown in Fig. 4. We shall refer to this object (in any of its three states) as a 'zipper'. As demonstrated in Fig. 3, this object can move through the defect-free RT phase, leaving topological reorganization in its path. The creation and annihilation of the zipper is achieved through transient defects that we shall discuss in the following Section.

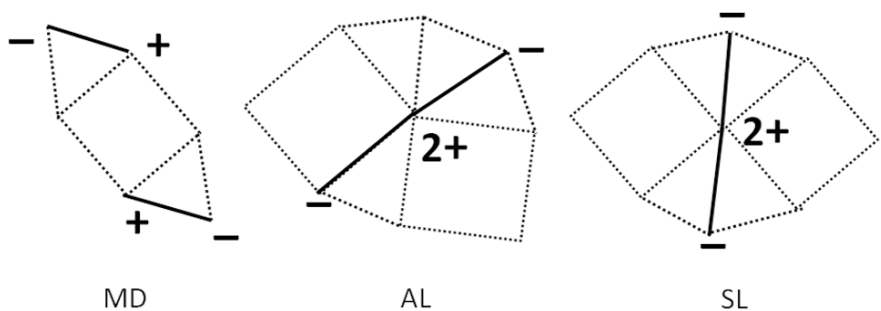

**Figure 4**. Three basic configurations of the zipper: the mobile doublet (MD), asymmetric line (AL) and symmetric line (SL). The strength of each disclination, as defined in Eq. 1, is indicated.

Despite the random arrangement of environments in the RT phase, there are similarities between the mobile 'zipper' observed in the RT simulations and the mobile defects associated with vacancies and interstitials in the single component triangular crystal. In both cases, the mobile defect consists of a pair of dislocations and propagates via transformations between three distinct configurations of these dislocations.

In the case of the zipper, these transformations occur via the elementary exchange between a square and adjacent triangle as depicted in Fig. 5. We shall refer to this elementary event as a flip. The set of possible 1-flip transformations of an isolated zipper are shown in Figs. 12-14 in Appendix I. With this information we can construct a diagram (shown in Fig. 6) depicting the connectivity of states of the zipper via a single flip.

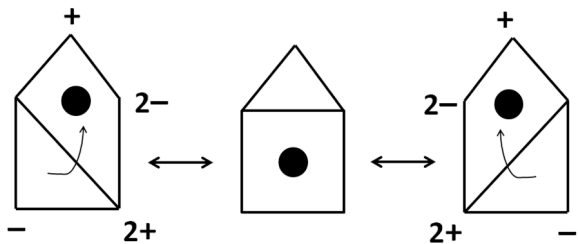

**Figure 5.** The two possible elementary flips starting from an initial square and adjacent triangle. The changes in the topological charge are indicated along with the associated movement of the small particle (filled circle) which must remain in the centre of the square.

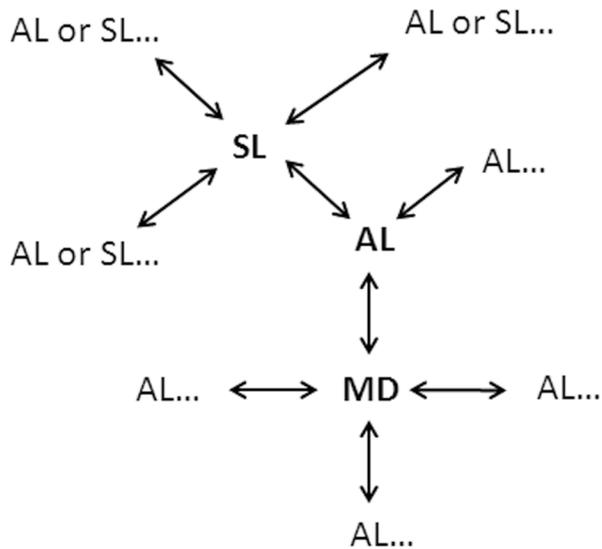

**Figure 6.** A schematic of the allowed transitions via single flips between the three zipper configurations: the mobile doublet (MD), asymmetric line (AL) and the symmetric line (SL). The shortest path between two distinct MD states involves two intermediate AL configurations.

## 3.2 Creation, Trapping and Annihilation of Zippers

While the zipper is propagated by the various sequences of flips depicted in Fig. 6, it cannot be annihilated nor rendered persistently immobile by action of these flips alone. This is a key feature of the mobile defect – its annihilation can only occur when it meets a SD defect. It follows, by application of microscopic reversibility, that the mobile defect is formed as a large defect object which subsequently divides (the reversal of the annihilation collision) to produce a mobile defect and an immobile defect.

The collision of the zipper with a SD defect (see Fig. 2) can result in three possible outcomes. The most common is that the zipper simply avoids close contact and continues on. If the angle of approach of the zipper corresponds to one specific angle, the two defects will mutually annihilate one another (an example of this process is shown in the movie provided in the Supplementary Material [31]). For another distinct angle of approach, the zipper and SD defect form a ST defect (see Figs. 2 and 7). The ST defect can persist as a stationary object for long time intervals before either reversing the capture and thus releasing the zipper to continue its journey or spontaneously annihilating itself via a concerted mechanism involving three internal flips (as shown in Fig. 7). All mechanisms of zipper creation observed to date result also in the formation of a SD defect as seen, for example, in the reversal of the process from Fig. 7b to Fig. 7a and in Fig. 3.

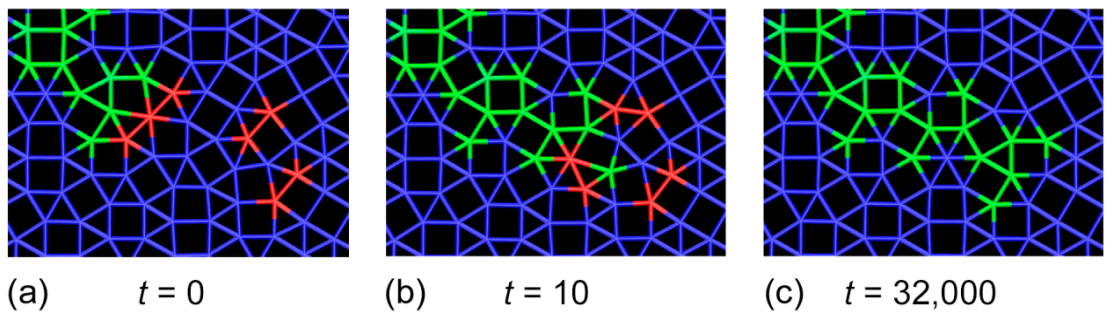

(a)     $t = 0$          (b)          $t = 10$          (c)     $t = 32,000$

**Figure 7.** Example of zipper capture and annihilation. In frame (a) we see the zipper on the left and the SD on the right. In frame (b), $10\tau$ later, the zipper has been captured and a static defect (the ST defect) is formed. In frame (c), $32,000\tau$ after frame (a), the ST defect has been annihilated via three concerted flips in the interior. The colour scheme is the same as in Fig. 3.

### 3.3 Stochastic Dynamics of the Zipper

The motion of the zipper corresponds to a stochastic walk on the connectivity diagram in Fig. 6. There are a variety of physical effects that render this walk non-Markovian and so distinct from a simple random walk: i) the flip is reversible so there is a significant probability of the zipper reversing previous steps; ii) the 12-fold orientational order breaks the circular symmetry of the plane; iii) the flip produces strain in the adjacent material which may bias the walk along preferred low strain directions; and iv) structural fluctuations in the form of domains consisting of all triangles or all squares may also present obstacles to the zipper. Maps showing the motion of the zipper's center of mass (COM) at $T = 0.45$ are presented in Fig. 8. The truncated side branches provide an unequivocal indication of the tendency for reversal and also the distance over which the reversal will persist in the form of short-range 'glide'-like behaviour.

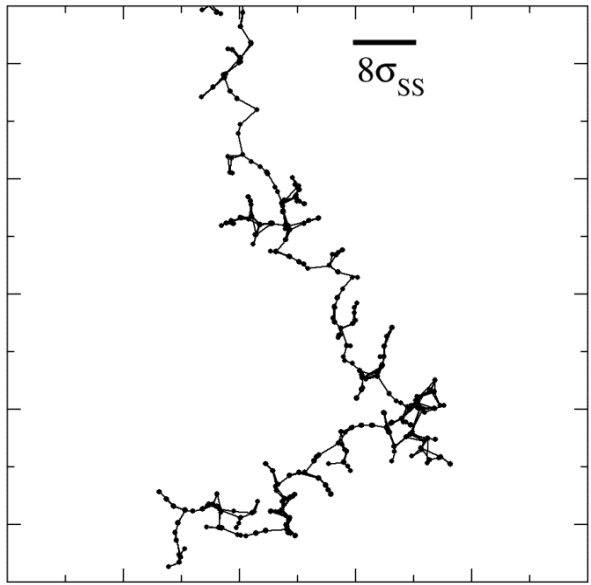

**Figure 8.** A trajectory of the zipper centre of mass (COM) at $T = 0.45$. Note the truncated side branches, characteristic of the spatial persistence of zipper reversals. The positions of the COM at $1\tau$ intervals are indicated by black dots. These clearly show that the zipper also has a strong tendency to persist moving in the same direction.

The distribution of the angle θ between sequential COM displacements, plotted in Fig. 9, provides one measure of how all of these influences combine to perturb the zipper motion from that of the random walk. To filter out the small displacements arising from round off error in the energy minimization (see Section 2) and to additional incorrectly identified defect vertices which skew the COM of the zipper, we have only registered a displacement when its magnitude exceeds a threshold length set to 0.25. The most striking deviation from the random walk behaviour seen in Fig. 9 is the large peak at θ = 180° (see inset) corresponding to a larger-than-random probability of reversal of the previous zipper move. More subtly, we find a peak at 0° – the vestige of the 'glide' that dominates defect motion in crystals – and additional peaks near 60° that appear to be associated with transitions between line defects (AL or SL, see Figs. 13 and 14). (The optimal fit for a small particle in a square of large particles with $\sigma_{LL}$=1.4 requires $\sigma_{SL}$=1.4/$\sqrt{2}$ ≈ 0.9899. If we choose this optimal value of $\sigma_{SL}$ then we find that the peak in Fig. 9 at 60° becomes considerably sharper.)

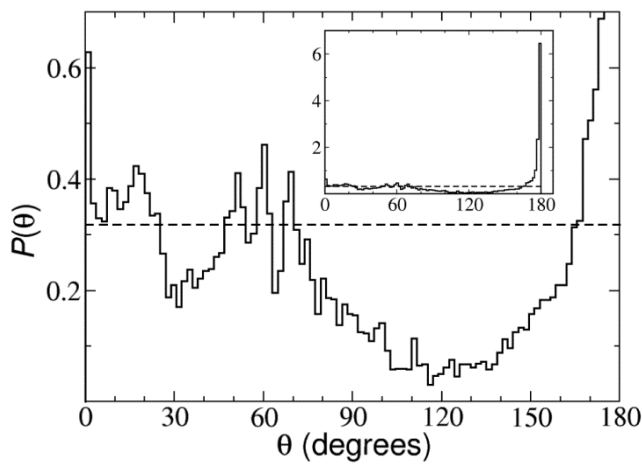

**Figure 9.** The distribution (at $T$ = 0.46) of the angle θ between subsequent displacements of the centre of mass (COM) of the zipper (defined as the COM of the constituent disclinations). The dashed line indicates the distribution for a random walk,

and the inset shows the full untruncated distribution. A threshold length of 0.25 was applied to the displacements to filter out noise (see text).

In Fig. 10 we plot the mean squared displacement (MSD) of the COM averaged over the choice of initial time. At long times the MSD increases linearly in time as expected for Fickian diffusion. The diffusion constant $D$ for the zipper, obtained from the asymptotic relation MSD$(t) = 4Dt$, ranges from $0.021 \pm 0.001$ at $T = 0.40$ to $0.062 \pm 0.003$ at $T = 0.48$, although we observe little variation above $T = 0.42$. At shorter times we observe sub-diffusive motion characterized by the relationship MSD $\sim t^b$ where $b < 1$. This type of behaviour is characteristic for a random walk subjected to obstacles or a non-random weighting of reversals, consistent with our observations from Fig. 9. For $T < 0.42$ the sub-diffusive behaviour persists out to a time scale of 1000 and a length scale of 10, while at $T \geq 0.42$ diffusion sets in around $t = 100$-$200$ and a length scale of $7.0 \pm 0.7$ (estimated from the time and length scale at which the crossover to diffusive behaviour occurs in the MSD plot). (Spatially coarse graining the trajectories until the reversal peak at 180° disappears suggests a crossover length that is closer to 12 at all $T$.)

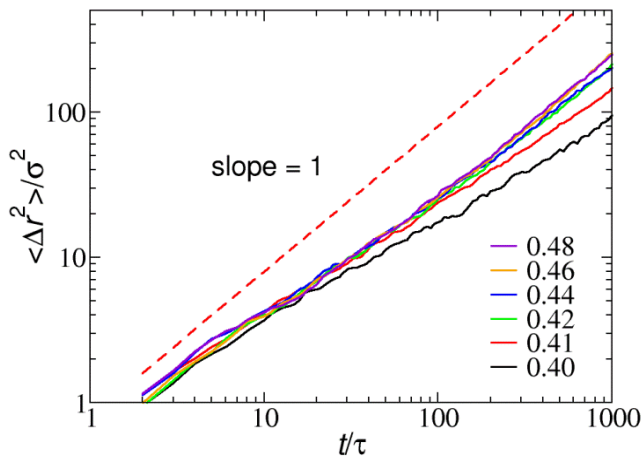

**Figure 10.** The mean squared displacement of the centre of mass of the zipper as a function of time for the range of temperatures indicated. The Fickian result with a slope of one is indicated by the dashed line.

### 3.4. Structural Relaxation via Zipper Motion

Having characterised the structure and dynamics of the zipper defect, we finally arrive at the central question of this paper – how effective is the diffusion of the zipper in relaxing the structure of the RT phase?

While we have presented a detailed analysis of the zipper structures and motion, we can only make the following general comment about the equilibrium density of zippers. The energetic cost of introducing a zipper and SD defect is $0.0003 \pm 0.0002$ at $T = 0.45$. This value is obtained as the difference in the average energy of configurations with and without zippers or the ST defects, all at $T = 0.45$. (Note that the zipper-less configurations do include short-lived thermal defects.) Based on this result, the zipper does not represent a significant energetic excitation. How large an equilibrium concentration this small excitation energy implies depends on the entropy gain provided by the zipper which, in turn, depends on the average lifetime of a zipper. We do not have sufficient statistics to estimate the zipper entropy and so can say little more about the equilibrium zipper concentration except to note that the low zipper concentration observed at $T = 0.45$ is likely to reflect, at least in part, the very slow equilibration time of the initial perfectly ordered configuration.

The relaxation function for the RT phase requires that we average over all choices of initial configurations – a requirement that will, at low temperatures and finite size samples, inevitably involve initial states without zippers. At low enough temperatures, the relaxation time will be dominated by the persistence time of the 'zipper-free' state – a timescale we cannot directly study. What we can address here is the time scale of structural relaxation in a configuration characterised by the presence of a single zipper. More specific to the question of defect-mediated relaxation, we shall compare the time dependence of structural relaxation with that of the fraction of atoms visited by the zipper. By this means, we separate the question of the effectiveness of an individual zipper to relax the structure around it from the more challenging problem of determining the equilibrium density of such defects.

Structural relaxation is typically defined as the relaxation of a Fourier component of density correlation between pairs of particles. This relaxation function, known as the *total* intermediate scattering function, is defined in Eq.2.

$$F(k,t) = \frac{1}{N} \langle \sum_{i=1}^{N} \sum_{j=1}^{N} \exp\left(i\boldsymbol{k} \cdot \left[\boldsymbol{r}_i(t) - \boldsymbol{r}_j(0)\right]\right) \rangle \qquad (2)$$

A variant of this relaxation function is the *self* intermediate scattering function which only considers the decay in correlation of individual particles with respect to their initial positions (see Eq. 3),

$$F_s(k,t) = \frac{1}{N} \langle \sum_{j=1}^{N} \exp\left(i\boldsymbol{k} \cdot \left[\boldsymbol{r}_j(t) - \boldsymbol{r}_j(0)\right]\right) \rangle \qquad (3)$$

In studies of relaxation in amorphous materials, the two relaxation functions are found to exhibit similar variation with time and temperature and, consequently, the self expression is often used as the sole monitor of relaxation as it is simpler to calculate and interpret. That the two intermediate scattering functions should behave similarly is by no means inevitable. Consider the following dynamic picture: a single atomic configuration whose dynamics consists of simply swapping particle labels between neighbouring pairs. The result is diffusion of individual particles without any change (and, hence, relaxation) of the original configuration. For such a model the self intermediate scattering function decays to zero while the total intermediate scattering function exhibits no time dependence at all.

In Fig. 11 we present the time dependence of the total and self intermediate scattering functions along with the fraction of particles that, at time $t$, are in the same topological environment as in the initial state. Comparing the total and self intermediate functions we find that, although there is some difference in the shapes of the functions at intermediate times, both functions decrease at a similar rate at long times. We conclude that the atomic rearrangements produced by the passage of the zipper through the RT phase result in structural relaxation. Next, we consider the relationship between structural relaxation and the dynamics of the zipper. Fig. 11 also shows that the time dependence of the fraction of topologically unchanged sites is very similar to that of the self intermediate scattering function. We therefore conclude that the statistics of the zipper motion is sufficient to describe structural relaxation of an RT configuration that includes a zipper. That said, our MD runs are not long enough to see complete structural relaxation and so it is possible that the preceding observation may require modification once the full relaxation curves are calculated.

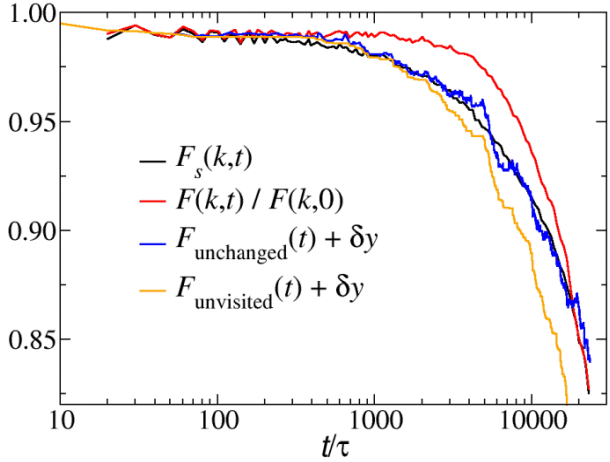

**Figure 11.** The total $F(k,t)$ and self $F_s(k,t)$ intermediate scattering functions for the RT phase (summing over large particles only) plotted alongside the fraction of particles $F_{unchanged}(t)$ that have not experienced a local topological change due to the action of the zipper after time $t$. For comparison we also show the fraction of particles $F_{unvisited}(t)$ that have yet to be visited by the zipper. $F(k,t)$ has been scaled by its value at $t = 0$ and $+\delta y$ indicates that a shift along the $y$-axis has been applied.

That the fraction of unchanged sites provides a good representation of the structural relaxation function provides strong support for a theory of structural relaxation based on defect diffusion. The previous work on defect diffusion [1-5] made use of the fraction of sites not visited by the defect rather than the fraction of sites left unchanged used in Fig. 11. This difference is due to the higher-than-random probability of reversals in the motion of the zipper. Such reversals undo the effect of the passage of the defect so that neglecting them, as is implicit in using the fraction of un-visited sites as a relaxation function, will lead to an underestimation of the structural relaxation time. The difference between these two measures of the zipper motion is shown in Fig. 11 where we compare the fraction of sites un-visited and the fraction of sites unchanged. As discussed, the former quantity clearly overestimates the rate of relaxation by roughly a factor of two.

This discussion of structural relaxation requires a qualification. The ideal RT phase exhibits a long range 12-fold orientational order, a feature omitted in the intermediate scattering function which is angularly averaged. While this order is reduced by including small strains due to the slight size mismatch of the lengths $\sigma_{LL}$ and $\sigma_{SL}$ used in this study, the zipper is not effective in relaxing this long range orientational order, a consequence of the peaks seen in the distribution of angles between zipper displacements shown in Fig. 9. We conclude this Section then with the qualification that the zipper is effective in relaxing that aspect of the structure described by the angular averaged intermediate scattering function but not, necessarily, all aspects of structure.

## 4. CONCLUSION

In this paper we have demonstrated how structural relaxation takes place at low temperatures in a dense phase characterised by a random assembly of 4 distinct local environments of the large particles in a binary mixture. We have been able to provide an explicit description of the structure of the diffusing defect, the mechanism of its movement and clear evidence of its capacity to relax structure in the RT phase. Where does that leave us? The square-triangle phase is a recurring feature of many quasicrystals and intermetallic compounds. The zipper defect we have identified relies only on the existence of disclinations in the tiling and so, potentially, represents a relaxation mechanism for planar square-triangle tilings stabilized by a variety of interactions. That said, the kinetic significance of the zipper will ultimately depend on the energy barriers associated with defect formation and motion in each specific case. Previous studies of diffusion in quasicrystals, discussed in Appendix II, have identified defect-free mechanisms for particle diffusion in which transitions between groundstate configurations can occur via small particle displacements not available to the particular atomic realization of the square-triangle phase studied here.

A diffusing defect mechanism of structural relaxation, along with the associated statistics of birth, annihilation and trapping, such as that provided here for the RT phase, provides an explicit account of the cooperative dynamics. Examples of Hamiltonian systems in which structural relaxation is achieved by diffusing defects, as in the binary mixture described here, are rare. The plaquette lattice model [32] is one of the more thoroughly studied examples of this behaviour. Whether the specific mechanism described here can also contribute to relaxation of a truly amorphous phase depends on the degree to which the higher concentration of static defects in glassy materials, on one hand, constrain zipper motion and lifetimes and, on the other, facilitate zipper creation. Examination of trajectories close to the melting point of the RT phase reveals that the zippers are readily generated at the surface of liquid-like drops within the RT phase and are rapidly propagated through the RT regions. (A movie showing zipper generation at the surface of a disordered droplet is provided in the Supplementary Material [31].) The possible extension of the relaxational role of zippers to the amorphous state is of interest as the defect-mediated relaxation mechanism represents a quite different approach [33], both qualitative and quantitative, to the general account of cooperative relaxation provided by the mosaic theory [34-36] that has been developed to describe relaxation near the glass transition in supercooled liquids. We believe that this paper provides a useful starting point to extending the study of diffusing defects and structural relaxation into phases characterised by increasing disorder.

**ACKNOWLEDGEMENTS**

The authors acknowledge a helpful discussion with Gilles Tarjus. The research has been supported by the School of Chemistry's Summer Scholar Program (E.L. and M.R.) and funding from the Australian Research Council through an Australian Postdoctoral Fellowship (A.W.) and a Discovery Project grant (P.H.).

**APPENDIX I**

**Single Flip Transformations Between the Zipper Configurations**

In Fig. 6 we sketched the connectivity of different configurations of the zipper via single flip events. In Figs. 12-14 below we present the explicit diagrams of the various possible single square-triangle flip transformations involving the three states of the zipper.

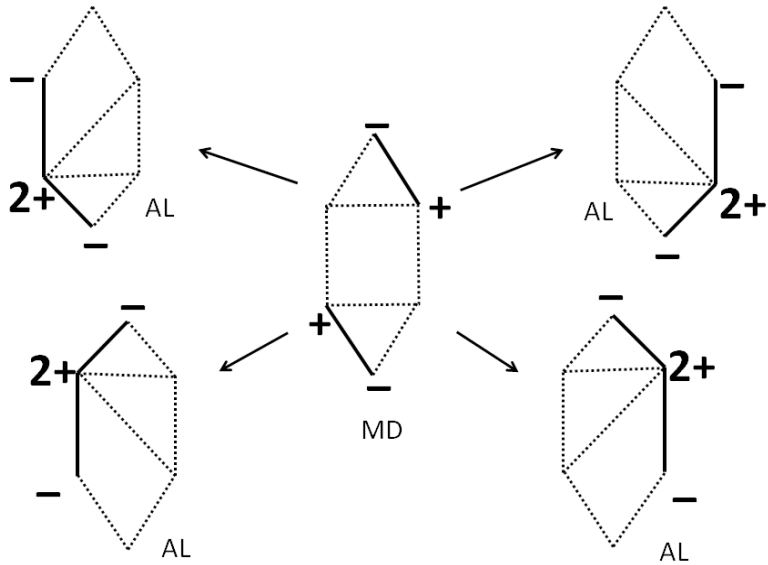

**Figure 12.** The four possible AL defects formed from a MD defect by a single flip. Topological charges are indicated.

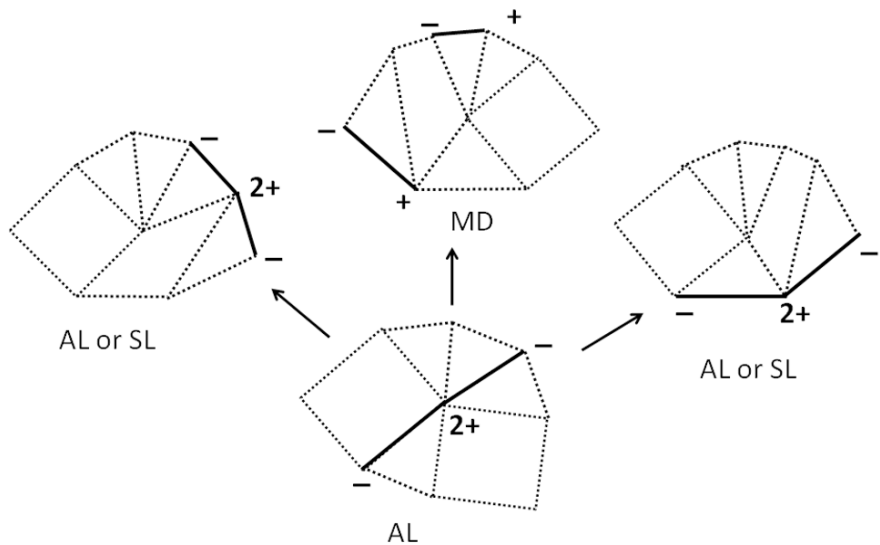

**Figure 13.** The three possible transitions from an AL via a single flip. Topological charges are indicated. The outcome 'AL or SL' depends on the geometry of the RT lattice beyond the nearest neighbor vertices of the original defect.

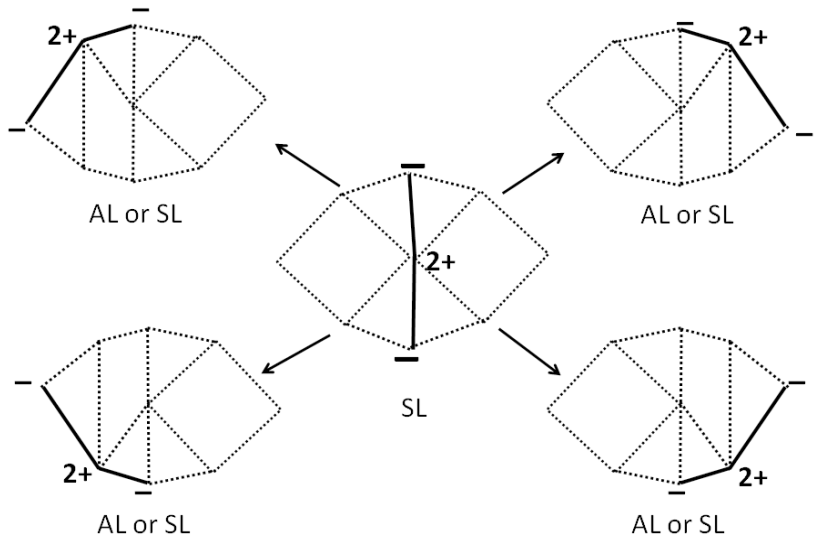

**Figure 14**. The four transitions from the SL configuration via a single flip. Topological charges are indicated. The outcome 'AL or SL' depends on the geometry of the RT lattice beyond the nearest neighbor vertices of the original defect.

**APPENDIX II**

**Related Studies of Defect-Mediated Diffusion in Quasicrystals**

The degeneracy of a quasicrystalline (or random tiling) configuration generally implies that a local reorganization of particles is sufficient to move between distinct ordered states. The localised rearrangements (not necessarily small ones) that connect quasicrystalline configurations have been expressed in the abstract notion of a *phason*. The entropic stabilization of a quasicrystal requires that the barrier for these localised rearrangements be small. It follows that these local events could contribute significantly to structural relaxation and diffusion (if the rearrangements can generate unbounded particle displacements) in the quasicrystal.

Phasons are typically introduced in terms of the orientation of the plane employed to project a higher dimensional periodic structure onto the real space dimension of interest. There have been a few studies that have considered explicit descriptions of the particle rearrangement in the real space. In this paper we have described how a defect, the zipper being the most elementary, is necessary for this transformation between ordered states in the case of the 2D binary mixture. There exist other quasicrystalline tilings in which defects are not required. Kalugin and Katz [37] proposed a transformation of an octagonal quasicrystal, related to a square-rhombus tiling, in which a small atom displacement, depicted in Fig. 15, changes the tiling without recourse to any defect.

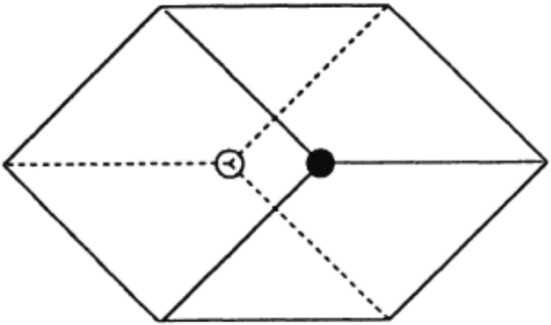

**Figure 15.** The shift of the central atom in an octagonal tiling of squares and rhombi between the open and filled circle positions.

The existence of a defect-free path between distinct ordered structures does not necessarily mean that defects do not play a significant role in the kinetics of particle diffusion. In Ref. 30 a binary atomic model of a dodecagonal quasicrystal was studied in which the tiling transformation, analogous to that suggested by Oxborrow and Henley [18] in the 2D tiling, involves only allowed configurations (see Fig. 16). In MD simulations of diffusion in this model Roth and Gähler [30] found that diffusion still involved the 'catalysis' of this transformation by defects such as vacancies.

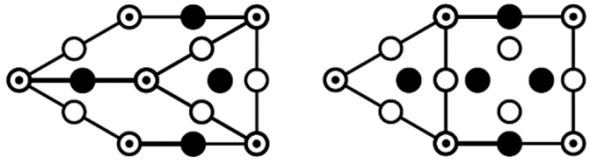

**Figure 16.** The transformation of a rhombus pair and a triangle to a triangle and square in the model of Ref. 30. This is a 3D model for which the 2D graphs represent the occupation of 4 distinct layers: white atoms are at z = 0, dotted atoms are at z = ¼ and ¾ and black atoms are at z = ½.